\documentclass[aps,prb,twocolumn,superscriptaddress,10pt]{revtex4-1}

\usepackage{graphicx}				
\usepackage{epsfig}
\usepackage{amsmath,amsthm} 
\usepackage{color}
\usepackage{verbatim}				
\usepackage{ulem}
\begin{document}

\title{Control of the asymmetric band structure
in Mn$_2$Au by a ferromagnetic driver layer}

\author{Y. Lytvynenko}
\affiliation{Institut f\"{u}r Physik, Johannes Gutenberg-Universit\"{a}t, Staudingerweg 7, D-55099 Mainz, Germany}
\affiliation{Institute of Magnetism of the NAS and MES of Ukraine, 03142 Kyiv, Ukraine}

\author{O. Fedchenko}
\affiliation{Institut f\"{u}r Physik, Johannes Gutenberg-Universit\"{a}t, Staudingerweg 7, D-55099 Mainz, Germany}

\author{S.V. Chernov}
\affiliation{Institut f\"{u}r Physik, Johannes Gutenberg-Universit\"{a}t, Staudingerweg 7, D-55099 Mainz, Germany}

\author{S. Babenkov}
\affiliation{Institut f\"{u}r Physik, Johannes Gutenberg-Universit\"{a}t, Staudingerweg 7, D-55099 Mainz, Germany}

\author{D. Vasilyev}
\affiliation{Institut f\"{u}r Physik, Johannes Gutenberg-Universit\"{a}t, Staudingerweg 7, D-55099 Mainz, Germany}

\author{O. Tkach}
\affiliation{Institut f\"{u}r Physik, Johannes Gutenberg-Universit\"{a}t, Staudingerweg 7, D-55099 Mainz, Germany}

\author{A. Gloskovskii}
\affiliation{Deutsches Elektronen-Synchrotron DESY, 22607 Hamburg, Germany}
\author{T.R.F. Peixoto}
\affiliation{Deutsches Elektronen-Synchrotron DESY, 22607 Hamburg, Germany}
\author{C. Schlueter}
\affiliation{Deutsches Elektronen-Synchrotron DESY, 22607 Hamburg, Germany}

\author{V. Grigorev}
\affiliation{Institut f\"{u}r Physik, Johannes Gutenberg-Universit\"{a}t, Staudingerweg 7, D-55099 Mainz, Germany}
\affiliation{Graduate School of Excellence Materials Science in Mainz, 55128 Mainz, Germany}
\affiliation{Department of Physics, AlbaNova University Center, Stockholm University, 10691 Stockholm, Sweden}

\author{M. Filianina}
\affiliation{Institut f\"{u}r Physik, Johannes Gutenberg-Universit\"{a}t, Staudingerweg 7, D-55099 Mainz, Germany}
\affiliation{Graduate School of Excellence Materials Science in Mainz, 55128 Mainz, Germany}
\affiliation{Department of Physics, AlbaNova University Center, Stockholm University, 10691 Stockholm, Sweden}

\author{S. Sobolev}
\affiliation{Institut f\"{u}r Physik, Johannes Gutenberg-Universit\"{a}t, Staudingerweg 7, D-55099 Mainz, Germany}

\author{A. Kleibert}
\affiliation{Paul Scherrer Institute, Swiss Light Source, 5232 Villigen, Switzerland}

\author{M. Kl{\"a}ui}
\affiliation{Institut f\"{u}r Physik, Johannes Gutenberg-Universit\"{a}t, Staudingerweg 7, D-55099 Mainz, Germany}

\author{J. Demsar}
\affiliation{Institut f\"{u}r Physik, Johannes Gutenberg-Universit\"{a}t, Staudingerweg 7, D-55099 Mainz, Germany}

\author{G. Sch{\"o}nhense}
\affiliation{Institut f\"{u}r Physik, Johannes Gutenberg-Universit\"{a}t, Staudingerweg 7, D-55099 Mainz, Germany}

\author{M. Jourdan}
\affiliation{Institut f\"{u}r Physik, Johannes Gutenberg-Universit\"{a}t, Staudingerweg 7, D-55099 Mainz, Germany}

\author{H.J. Elmers}\email{elmers@uni-mainz.de}
\affiliation{Institut f\"{u}r Physik, Johannes Gutenberg-Universit\"{a}t, Staudingerweg 7, D-55099 Mainz, Germany}

\keywords{}

\date{\today}

\begin{abstract}
Hard X-ray angle-resolved photoemission spectroscopy reveals the momentum-resolved band structure in an epitaxial Mn$_2$Au(001) film capped by a 2~nm thick ferromagnetic Permalloy layer. By magnetizing the Permalloy capping layer, the exceptionally strong exchange bias aligns the N{\'e}el vector in the Mn$_2$Au(001) film accordingly. Uncompensated interface Mn magnetic moments in Mn$_2$Au were identified as the origin of the exchange bias using X-ray magnetic circular dichroism in combination with photoelectron emission microscopy.
Using time-of-flight momentum microscopy, we measure the asymmetry of the band structure, $E(k)\neq E(-k)$, in Mn$_2$Au resulting from the homogeneous orientation of the N{\'e}el vector. Comparison with theory shows that the  N{\'e}el vector, determined by the magnetic moment of the top Mn layer, is antiparallel to the Permalloy magnetization. The experimental results 
demonstrate that hard X-ray photoemission spectroscopy can measure the band structure of epitaxial layers beneath a metallic capping layer
and corroborate the asymmetric band structure in Mn$_2$Au that was previously inferred only indirectly.
\end{abstract}
\maketitle

\section{Introduction} 
Antiferromagnets have attracted scientific interest as active materials in 
spintronics~\cite{Jungwirth2016, Gomonay2017,Baltz2018}.
The orientation of the sublattice magnetization, defined as the N{\'e}el vector $\vec{N}=\vec{M_1}-\vec{M_2}$, acts as an information carrier, where
the anisotropic magnetoresistance effect allows to read out 
the N{\'e}el vector alignment~\cite{Reimers2023}.
Magnetic linear dichroism allows for an ultrafast optical read-out~\cite{Grigorev2021}. 
Electrical currents enable the manipulation of $\vec{N}$ to write information, as has been demonstrated for 
CuMnAs~\cite{Zelezny2014,Wadley2016,Grzybowski2017,Wadley2018} and Mn$_2$Au~\cite{Bodnar2018,Meinert2018,Reimers2023}.  
A second-order magnetoresistance has been observed in CuMnAs~\cite{Godinho2018}.
For Mn$_2$Au films spin to charge current conversion by the inverse spin Hall effect
has been shown~\cite{Arana2018} as well as the
electric field control of the N{\'e}el spin–orbit torque~\cite{Chen2019}.
The current-induced manipulation of $\vec{N}$ has been attributed to
the N{\'e}el spin-orbit torque~\cite{Gomonay2017}.

At the microscopic electronic-structure level, 
the control over the N{\'e}el vector allows manipulation of some electronic properties~\cite{Linn2023}.
The manipulation of electronic states in Mn$_2$Au
shows up as an asymmetry in the electronic band structure, $E(k) \neq E(-k)$~\cite{Fedchenko2022}.
The usual band structure parity $E(k) = E(-k)$ results from the symmetries of the materials, such as inversion/parity $\mathcal{P}$, time-reversal coupled with translation $\mathcal{T}\boldsymbol{t}$, or
time-reversal coupled with the spin rotational symmetry $\mathcal{TR}_{S}$
($\mathcal{R}_S$ rotates the spin by 180$^0$). 
In Mn$_2$Au these symmetries are broken.
The asymmetric band structure is then caused by
a Rashba-like mechanism in combination with the staggered magnetization in Mn$_2$Au~\cite{Smejkal2017,Smejkal2017b}. 



Predictions for the occurrence of antiferromagnetic parity violation 
have been given for spin–orbit coupled collinear antiferromagnets such as CuMnAs and Mn$_2$Au, which break $\mathcal{P}$ and $\mathcal{T}$ symmetries but preserve the combined $\mathcal{PT}$ 
symmetry~\cite{Smejkal2017,Hayami2018,Watanabe2020,Elmers2020,Hayami2020}. 
Despite having crystal parity, the antiferromagnetic order
breaks this symmetry in the presence of spin-orbit coupling.
The direct observation of parity violation in Mn$_2$Au using 
angle- and space-resolved photoemission spectroscopy (ARPES) was reported in Refs.~\onlinecite{Fedchenko2022}.

The occurrence of band asymmetry is not restricted to these two systems but also
shows up at interfaces of collinear antiferromagnets, as reported for  GdIr$_2$Si$_2$~\cite{Schulz2021}, and for Ag$_2$Bi-terminated Ag films~\cite{Carbone2016}. 
Asymmetric band structures have also been predicted to occur in complex non-coplanar magnets that break the $\mathcal{TR}_{S}$ symmetry~\cite{Hayami2018}.
However, in these systems the N{\'e}el vector is determined by the crystal structure
and cannot be manipulated by current.

The presence of small antiferromagnetic domains in Mn$_2$Au~\cite{Sapozhnik2018,Bommanaboyena2021} is generally an obstacle to angle-resolved photoemission spectroscopy, which then requires both angular and spatial resolution~\cite{Fedchenko2022}. 
As we have shown previously~\cite{Bommanaboyena2021}, it is possible
to align the N{\'e}el vector using 
a ferromagnetic Permalloy driver layer deposited on top of the
Mn$_2$Au film by exploiting the strong exchange bias between Mn$_2$Au and Permalloy. 
The remaining questions are what is the relative orientation of the N{\'e}el vector and the magnetization in the Permalloy driver layer,
and can one observe the controlling of the band structure in Mn$_2$Au
beneath the driver layer.

In this article, 
we confirm the control of the asymmetric band structure in a Mn$_2$Au(001) film by the remanent magnetization in the Permalloy layer. 
Our results show an antiparallel orientation between the magnetic moments
in the Permalloy layer and 
in the top Mn layer, indicating the  N{\'e}el vector direction in the Mn$_2$Au film.
We use hard X-ray angular resolved photoemission spectroscopy
to determine the electronic structure of Mn$_2$Au(001) films below the
capping layer utilizing the increased inelastic mean free path of high-energy photo-excited electrons. 
Here, the Permalloy layer is polycrystalline and therefore only adds an angular-independent photoemission intensity to the background.

\section{Experimental Details}
All layers of the Mn$_2$Au(001)(45~nm)/Ni$_{80}$Fe$_{20}$(Py)
(2~nm)/SiN$_x$(1.8~nm) samples were deposited by RF sputtering on epitaxial Ta(001)(13~nm) single or Ta(001)(13~nm)/Mo(001)(20~nm) double buffer layers on the Al$_2$O$_3$(r-plane) or MgO(100) substrates, respectively. The details of the deposition process are described in Refs.~\onlinecite{Jourdan2015,Satya2020,Elmers2020}. The Permalloy and SiN$_x$ layers were deposited at room temperature and form polycrystalline films. SiN$_x$ is an oxidation-preventing capping layer, and is highly transparent to photoemitted electrons~\cite{Elmers2020}.
Rectangular magnetic hysteresis loops of the samples, measured using a Quantum
Design MPMS SQUID-magnetometer, confirm that the saturation magnetization 
in Permalloy is equal to the remanent magnetization (square-shaped hysteresis loop). Magnetic alignment of Mn$_2$Au(001)/Py samples was performed in a magnetic field 
up to 1~T applied in four different easy crystallographic directions 
$<110>$ of Mn$_2$Au.
For some samples, the Permalloy layer was not added 
to allow for reference measurements in the as grown
multidomain state of the antiferromagnetic Mn$_2$Au film.

Hard X-ray photoemission spectroscopy was performed at beamline P22
of the storage ring PETRA III at DESY in Hamburg
 using time-of-flight momentum microscopy~\cite{Medjanik2017}. 
Due to the high energy (6.0~GeV) and the
large size (2.3~km circumference) of PETRA, P22 provides
hard X-ray radiation with high brilliance
in an energy range from 2.4 to 15~keV. 
The experimental conditions were 2$\times 10^{13}$ photons/s at 5.210~keV in a spot of about
50$\times 50$~$\mu$m$^2$ using a Si(311) double-crystal monochromator~\cite{Medjanik2019}. 
The total energy resolution is
governed by the photon band width of 155~meV.

For hard X-ray measurements with defined orientation of
the N{\'e}el vector direction, 
we cut each 1$\times 1$~cm$^2$ film into four pieces, 5$\times 5$~mm$^2$ each.
We then mounted the four pieces on the same sample holder, 
where for each piece the Permalloy layer was magnetized prior to mounting in one of the four easy magnetic directions of Mn$_2$Au along the $<110>$ crystal axes. 
Photoemission data were then recorded separately for each of the four mounted pieces.

Similar experiments were carried out on samples with different thicknesses of Permalloy. Using a thinner 1.5~nm thick Permalloy layer results in a better signal-to-background ratio.
However, the hysteresis loop is no longer square-shaped and the remanent magnetization is significantly reduced. 
A thicker (3~nm) Permalloy layer exhibits a larger coercive field, but the band structure is barely visible due to the low signal-to-background ratio, 
which would require unrealistically long data acquisition times.

Photoemission electron microscopy measurements (PEEM) on SiN-capped Mn$_2$Au films were performed at the SIM beamline of the Swiss Light Source. The sample was illuminated by circularly polarized X-rays at a grazing angle of 16$^0$. 
The X-ray magnetic circular dichroism (XMCD) image was then calculated as the normalized difference between the images recorded with right and left circularly polarized light. 
To plot the XMCD contrast as a function of the incident photon energy, we calculated the normalized difference between the mean value within the black and white domains
(see Fig.~\ref{FigXMCD}) over the entire field of view.

\section{Results}

\begin{figure}
\includegraphics[width=\columnwidth]{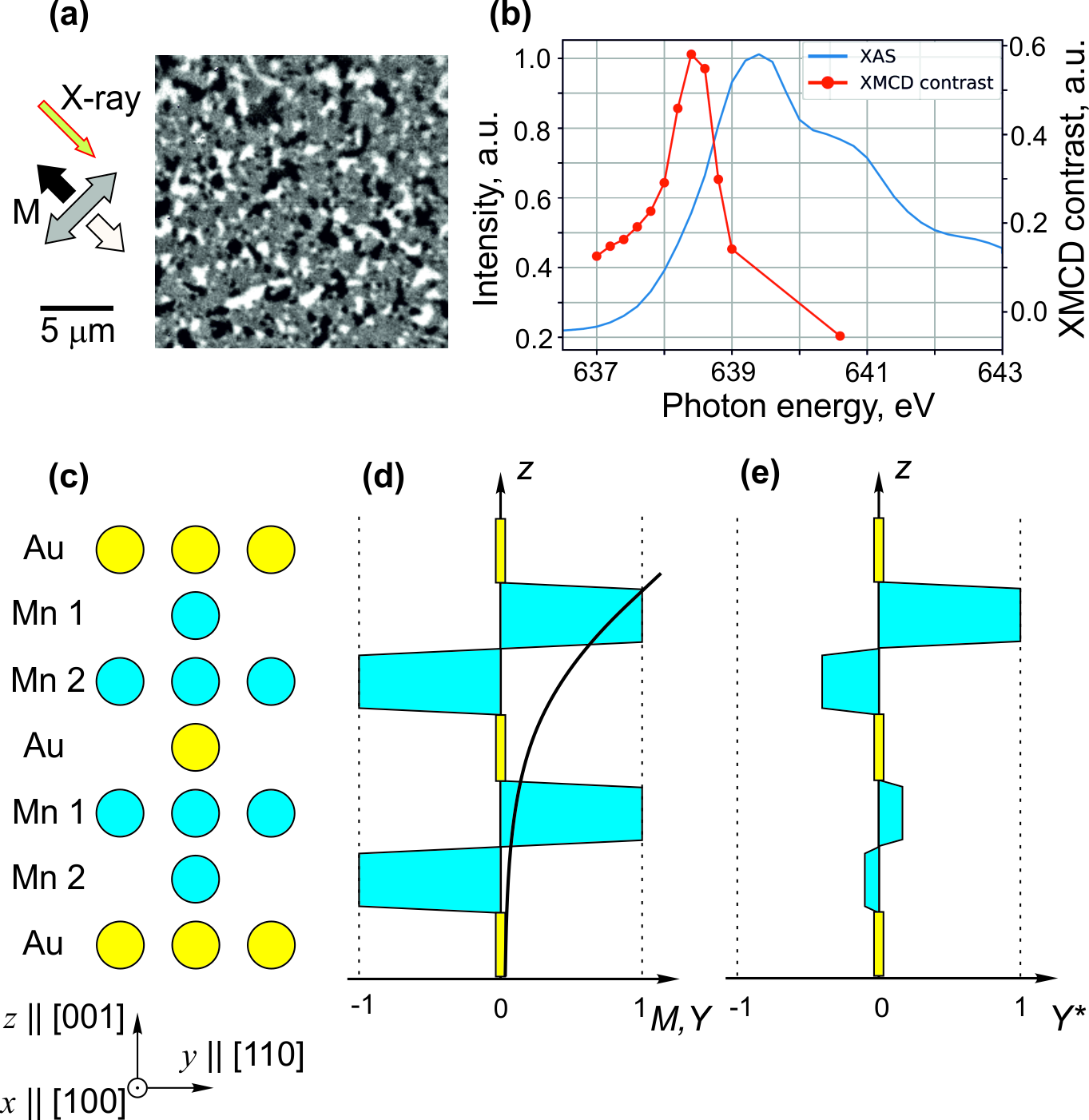}
\caption{\label{FigXMCD} 
(a) XMCD-PEEM image of the Mn$_2$Au(100) film capped with a thin SiN layer
measured at the photon energy of 638.5~eV.
Black and white contrast levels indicate the XMCD contrast of opposite sign, whereas the intermediate contrast level indicates a vanishing XMCD value,
corresponding to the domains with N{\'e}el vector aligned perpendicular to the X-ray propagation direction (see schematic to the left of the XMCD contrast image).
(b) X-ray absorption spectrum (blue full line) and XMCD signal (red dots) {\it versus} photon energy.
(c) Sketch of the layered crystal structure of Mn$_2$Au.
(d) Sketch of the electron yield $Y$ for an atomic layer and the atomic layer magnetization $M$ versus depth $z$. 
(e) Resulting contribution $Y^*$ of the atomic layers to the total XMCD signal,
showing that the prevailing component stems from the top Mn layer.
}
\end{figure}

\begin{figure*}
\includegraphics[width=\textwidth]{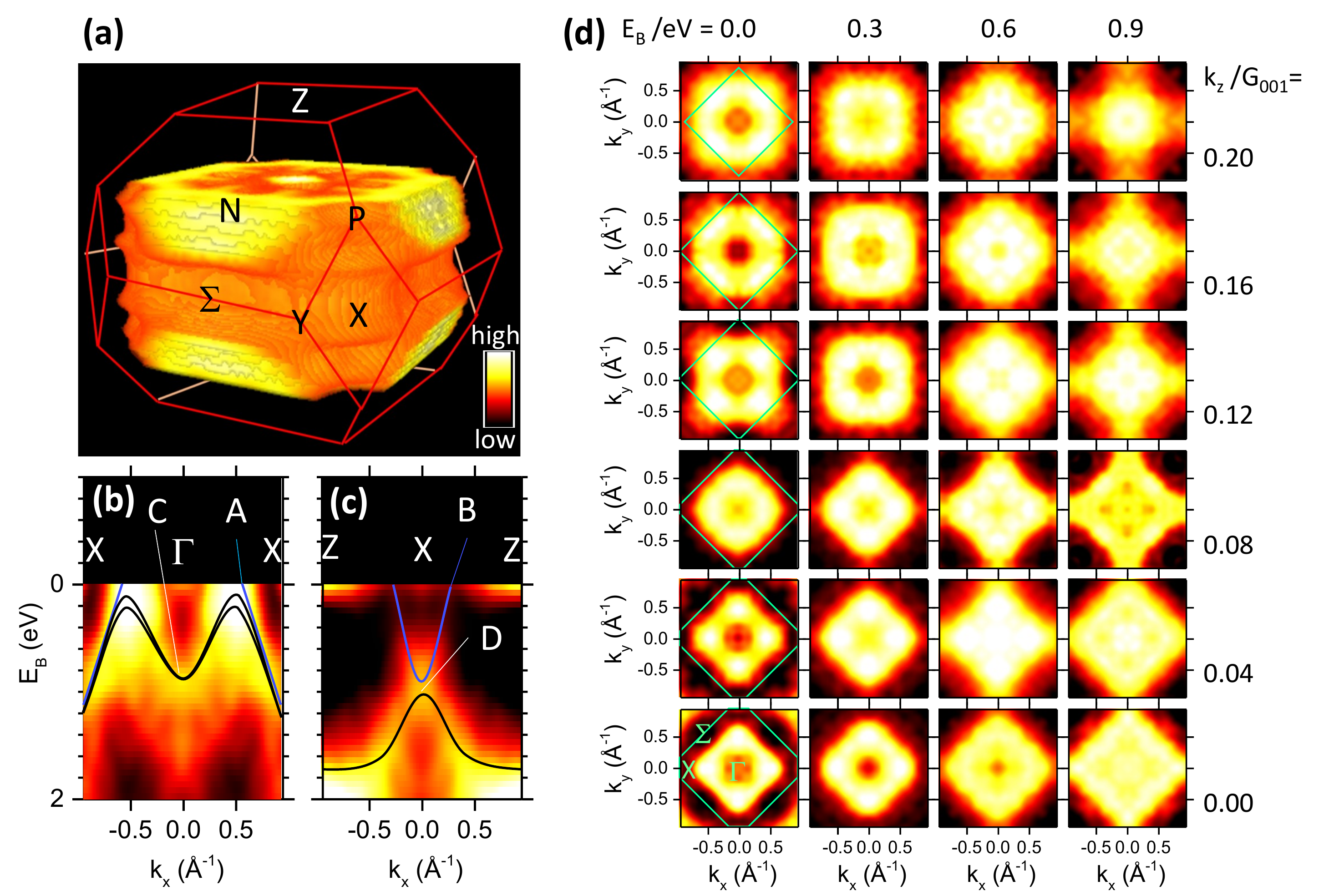}
\caption{\label{Fig1} 
(a) Three-dimensional representation of the measured
spectral density of states of Mn$_2$Au at the Fermi level
$I(E_B=0,k_x,k_y,k_z)$ (Fermi surface) for $-0.25<k_z/G_{001}<0.25$.
The indicated Brillouin zone (red lines) of the body centred tetragonal (bct$_2$) structure of Mn$_2$Au is based on Ref.~\onlinecite{Setywan2010}.
The data were extracted from hard X-ray
photoemission excited by a photon energy of 5210~eV at a
temperature of 20~K.
(b,c) Band dispersions along the X-$\Gamma$-X and Z-X-Z direction, respectively.
The color code represents the photoemission intensity from black to white (red-hot)
after background subtraction. Full lines denote calculated bands from 
Ref.~\onlinecite{Elmers2020}.
(d) Constant-energy maps of the spectral density of electronic states
$I(E_B,k_x,k_y,k_z)$ in planes perpendicular to the c-axis.
Note that the square areas extend to the Z-Y-$\Sigma$ plane of the adjacent Brillouin zones towards the corners (Z-points) of the images.
Perpendicular momentum $k_z$ values are given with respect to the 
nearest $\Gamma$-point.
}
\end{figure*}

\begin{figure}
\includegraphics[width=\columnwidth]{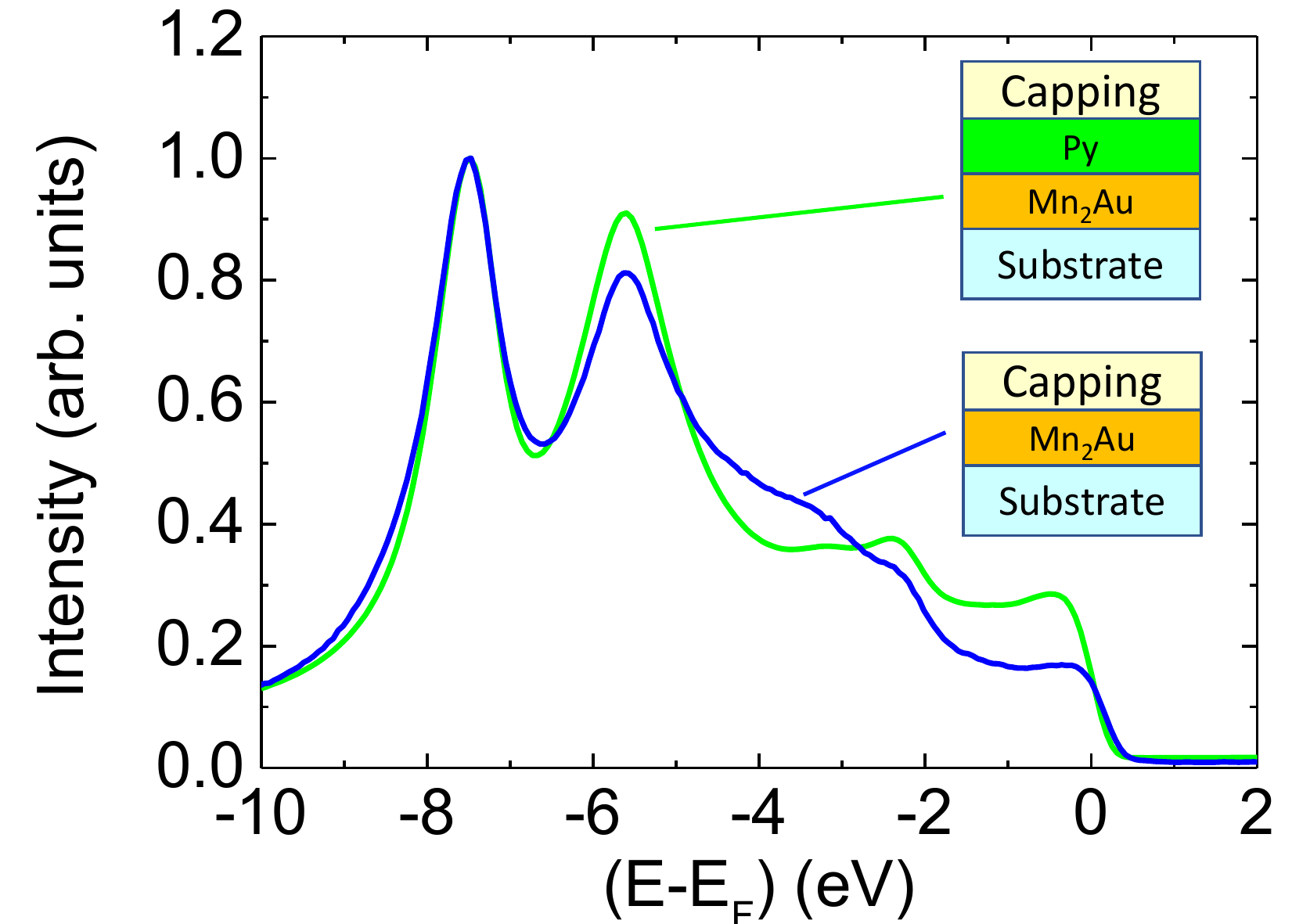}
\caption{\label{Fig2} 
Energy distribution curves obtained from integrating the photoemission intensity over the parallel momentum for a photon energy of 5210~eV
for the sample without Permalloy layer and 5185~eV for the
sample with Permalloy capping.
The binding energy is $E_B=-(E-E_F)$.
The intensities are normalized at the peak near $E_B=8$~eV}
\end{figure}

\begin{figure*}
\includegraphics[width=0.7\textwidth]{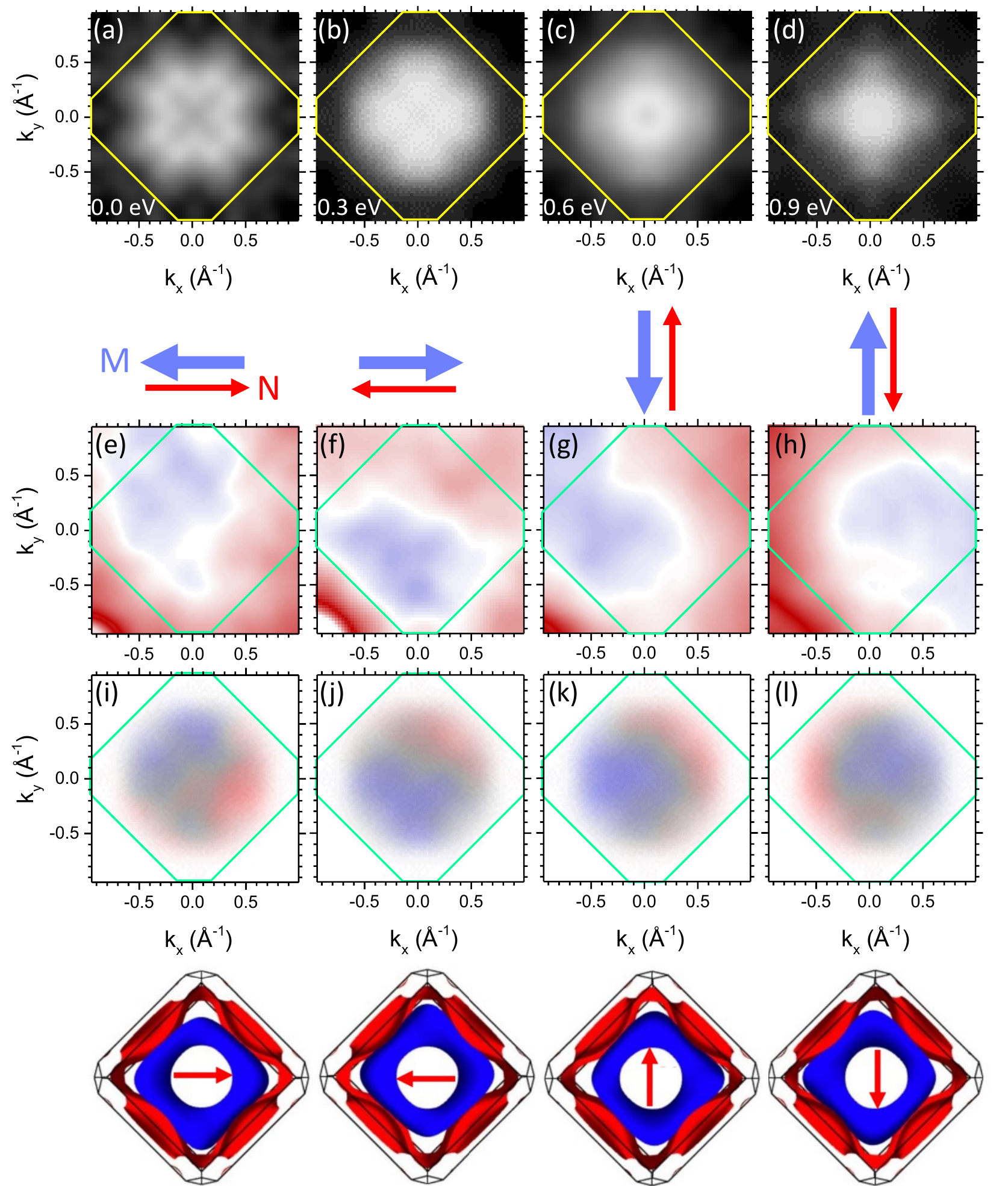}
\caption{\label{Fig3} 
(a-d) Symmetrized constant energy maps of the intensity distribution for Mn$_2$Au(100) capped with 2~nm Permalloy (Py) and 2~nm SiN protective layer measured at a photon energy of 5185~eV. The binding energy increases from (a) 0~eV ($E_F$), (b) 0.3~eV, (c) 0.6~eV to (d) 0.9~eV. Yellow lines indicate the Brillouin zone cut in the $\Gamma$-X-$\Sigma$ plane.
(e-h) Intensity difference maps $D(k_x,k_y) = I(E_B=0.5$~eV$,k_y,k_y)- I(E_B=0.2$~eV$,k_y,k_y)$ for the indicated magnetization orientation (M) in the Py layer and the corresponding N{\'e}el vector orientation (N). Red/blue color indicates positive/negative difference values.
(i-l) Difference and intensity depicted in a combined color scale.
The difference in experimental intensities stems from integration over energy and  momentum ranges given by the experimental resolution.
Bottom row: Calculated constant energy
surfaces in momentum space at $E_B=0.4$~eV for the indicated N{\'e}el vectors $N$ (red arrows). The red surface relates to band A [Fig.~\ref{Fig1}(b)] and the blue asymmetric surface to band C. (Calculated data from Ref.~\onlinecite{Fedchenko2022}).
}
\end{figure*}

A prerequisite for the exceptionally strong exchange bias field required to orient the N{\'e}el vector in the antiferromagnetic layer,
is the presence of uncompensated magnetic moments 
of Mn$_2$Au at its interface with the ferromagnetic layer.
The presence of uncompensated moments has been inferred indirectly 
from the well-defined morphology of the Mn$_2$Au surface,
which has been shown to display step heights of three atomic layers~\cite{Bommanaboyena2021}.
Here, we directly probe the Mn surface magnetization using X-ray magnetic circular dichroism (XMCD). 

Fig.~\ref{FigXMCD}(a) shows a photoelectron emission microscopy (PEEM) image
of the Mn$_2$Au(100) surface covered by a thin SiN layer,
where the XMCD contrast reveals the presence of antiferromagnetic domains.
The incident X-ray beam, which defines the quantization axis parallel to the magnetic easy axis of the antiferromagnetic film, arrives from the left with an angle of incidence of 74 degrees with respect to the surface normal. 
The contrast image shows three distinct contrast levels, which are expected if the N{\'e}el vector is parallel, antiparallel or perpendicular to the 
in-plane component of the X-ray photon momentum,
analogous to the case of a ferromagnet with fourfold magnetic anisotropy. 
The spectral information shown in Fig.~\ref{FigXMCD}(b) confirms the previously observed behavior~\cite{Sapozhnik2017} of metallic Mn in Mn$_2$Au and the pronounced
XMCD signal on the rising edge of the Mn L$_3$ absorption maximum.

The next step is to understand why the XMCD signal is non-zero
in view of the collinear antiferromagnetism in Mn$_2$Au.
To explain this phenomenon, we consider the exponentially 
decreasing electron yield from an atomic layer at a depth $z$ below the surface,
as depicted in Fig.~\ref{FigXMCD}(c-e). 
Because of this exponentially decreasing contribution with increasing depth,
the XMCD signal is dominated by the magnetization of the top Mn layer [Fig.~\ref{FigXMCD}(e)].
In combination with the surface morphology of the Mn$_2$Au(100) film,
which has step heights of precisely three atomic layers~\cite{Bommanaboyena2021}, 
the top Mn layer has the same magnetization orientation  on adjacent terraces in the case of an antiferromagnetic domain with homogeneous N{\'e}el vector orientation (see Fig.~\ref{FigXMCD}).

The surface morphology in combination with the collinear antiferromagnetism
in Mn$_2$Au thus explains the occurrence of the XMCD signal 
over domains much larger than the typical terrace width of epitaxially grown thin films.
The domain sizes observed here are consistent with previously reported 
domain sizes in epitaxial Mn$_2$Au films~\cite{Sapozhnik2018}.
The top Mn layer thus represents the largest possible surface density of uncompensated moments,
which then result in the strong exchange bias when the Mn$_2$Au 
is covered by a Permalloy layer.

We now turn to the hard X-ray photoemission experiments.
First, we report hard X-ray angular photoemission spectroscopy results on Mn$_2$Au films
where the Permalloy film has not been added, providing a reference for the Permalloy capped films. 
In this case the films exhibit a multi-domain antiferromagnetic state resulting in a symmetric band structure corresponding to the crystal symmetry.
We exploit the 4-fold symmetry around the c-axis to increase the signal-to-noise ratio $\it via$ symmetrization.

Fig.~\ref{Fig1}(a) shows the Fermi surface of Mn$_2$Au in
a three-dimensional color-coded intensity map.
Data for different $k_z$ values are derived from
repeated Brillouin zones measured simultaneously:
The large momentum field of view of 12~\AA$^{-1}$ includes five adjacent
Brillouin zones, where increasing parallel momentum
results in decreasing perpendicular momentum. 
This allows us to measure the perpendicular momentum $k_z$
in one experimental run in a range of $\Delta k_z = 0.5G_{001}$~\cite{Agustsson2021}.
The $k_z$ interval where photoemission intensity above the background level could be detected
is actually smaller ($\Delta k_z = 0.25G_{001}$). 
This is likely due to a 
destructive photoelectron interference~\cite{Schonhense2019}.

The experimentally obtained Fermi surface is in qualitative agreement with the theoretically predicted Fermi surface reported in Ref.~\onlinecite{Elmers2020}. 
In particular, the diamond-shaped cross section in 
the $\Gamma$-X-$\Sigma$ plane
agrees well with calculations. 
Furthermore, the high intensity observed near the N-points corresponds
to the calculated tubular bands crossing the N-Y-P plane.

The band dispersions shown in Fig.~\ref{Fig1}(b,c) 
are derived from slices through the 4-dimensional array 
$I(E_B,k_x,k_y,k_z)$, where the $k_z$-value has been determined with a better accuracy as compared to the experimental results reported in Ref.~\onlinecite{Elmers2020}.

The inverted parabolic band A, shown in 
Fig.~\ref{Fig1}(b) along the $\Gamma$-X profile,
has an apex above the Fermi level. 
The band crosses the Fermi level
near 0.5~\AA$^{-1}$ and forms the Fermi surface.
The band C, which is below band A, disperses to a larger binding energy as it approaches the $\Gamma$-point, forming maxima near $k_x = \pm 0.5$~\AA$^{-1}$ and a minimum at $\Gamma$. 
This central minimum of band C appears at
$E_B = 0.9$~eV, in very good agreement with previously reported results~\cite{Elmers2020}, 
and deviates from the calculated
binding energy of $E_B = 0.5$~eV, revealing a shift of 400~meV to higher binding energies, 
due to electron correlation effects not included in the theoretical model. 
The energy broadening of band C near $k_x = \pm 0.5$~\AA$^{-1}$
is associated with the splitting of band C at $|k_x| > \pm 0.4$~\AA$^{-1}~\cite{Elmers2020}$, 
which is a result of the
spin-orbit coupling and thus depends on the orientation of the N{\'e}el vector.
Note that the averaging over magnetic domains prevents the observation
of the band  asymmetries in this configuration.

The $E_B$ {\it versus} $k_{x}$ section along the Z-X-Z direction
[Fig.~\ref{Fig1}(c)] shows a prominent band D near the Z point at $E_B = 1.7$~eV and at X at  $E_B = 1$~eV. 
In this case the binding energy is in agreement
with the calculated results~\cite{Elmers2020}. 
The broadening of the band D observed for $|k_x| > \pm 0.4$~\AA$^{-1}$ 
is again related to the spin-orbit coupling. 
The parabolic band B shows an apex at the X-point at $E_B$ = 1~eV. 
For band B we find a good agreement with the calculated values 
given in Ref.~\onlinecite{Elmers2020}.
The  band gap between bands B and D at this position
depends on the N{\'e}el vector orientation and vanishes in the 
domain averaged data used here. 
The photoemission intensity at the Z-points near the Fermi level 
is probably associated with bands forming the Fermi surface in the Z-$\Sigma$-Y plane,
which appear with weak intensity due to a small photoemission matrix-element~\cite{Fedchenko2022}.

Fig.~\ref{Fig1}(d) shows sections through the 4-dimensional data set
$I(E_B,k_x,k_y,k_z)$ at the given perpendicular moments $k_z$ and
binding energies $E_B$. This plot serves as a reference
for the photoemission results obtained on N{\'e}el vector aligned Mn$_2$Au films capped with
the ferromagnetic Permalloy layer.


The energy distribution curves integrated over the parallel momentum
at $k_z=0$, shown in Fig.~\ref{Fig2}, compare films with and without
Permalloy layer capping. The prominent peaks near the binding energies
$E_B=6$ and 8~eV are due to the localized Au 5$d$ states in the 
Mn$_2$Au film~\cite{Zuazo2007}. 
Note that the high photon energy causes an increase of the photoemission probability with increasing orbital momentum. 
The energy distribution shows that the Au 5$d$ peaks are almost unchanged
after the capping with Permalloy, confirming the increased inelastic mean free path at this high photon energy.
The photoemission intensity for the capped sample is increased
in the range $E_B<2$~eV. 
This corresponds to the broad photoemission intensity maximum 
that is observed for bulk Permalloy in this energy range~\cite{Altmann1999}.
It is therefore expected that the photoemission intensity near the Fermi level originating from Mn$_2$Au is enhanced by an unstructured background intensity
photo-excited from the Permalloy layer. 
The top SiN layer, which prevents oxidation of the film stack,
has an energy gap in this binding energy range and therefore
does not lead to additional photo-electrons.


Figs.~\ref{Fig3} (a-d) show the momentum-resolved
photoemission intensity for the sample with the Permalloy layer.
The increase of the elastic background intensity without momentum
information is twofold compared to the uncapped Mn$_2$Au film
because of the additional photoemission yield from the Permalloy layer.
In addition, the probability that photoelectrons excited
in the Mn$_2$Au layer are scattered by phonons in the 
Permalloy layer during the motion towards the surface adds up to the Debye-Waller scattering
occurring in the   Mn$_2$Au layer itself.
Due to the limited energy resolution in the hard X-ray regime,
these quasi-elastic electrons also add up to the momentum-independent background intensity. Hence, the background intensity relative to the direct photoemission intensity is significantly smaller than in the case of films without a Permalloy layer.

To increase the signal-to-background ratio, we symmetrized the constant energy maps 
according to the 4-fold crystal symmetry about the c-axis.
In principle, the homogeneous background intensity can be subtracted
from the measured data. 
However, the signal-to-noise ratio is determined by the total measured intensity.
Therefore, increasing the background intensity about fourfold,
as in the case of the 2~nm Permalloy capping,
would require an increase in acquisition time by a factor of 16 to 
achieve the same data quality or to sacrifice momentum resolution by binning. 
This explains the decreased effective momentum resolution in 
Figs.~\ref{Fig3}(a-d).

The symmetrized results shown in Fig.~\ref{Fig3}(a-d) confirm that the capping with 
Permalloy still allows the band structure of the epitaxial Mn$_2$Au film  beneath the Permalloy layer to be observed.   
The sequence of constant energy maps for binding energies between 0 and 0.9~eV
for a photon energy of 5185~eV
agrees with the reference data for similar films without the Permalloy capping for
$k_z/G_{001}=0.1$ (see Fig.~\ref{Fig1}). 
The deviation from $k_z/G_{001}=0.0$ is due to the slightly lower photon energy
set for this experiment.
Therefore, the data confirms that the Mn$_2$Au band structure can be detected beneath the Permalloy film.

Finally, we discuss the non-symmetrized data for the samples with
different Permalloy magnetization directions.
We subtract the photoemission intensity recorded at two different binding energies 
for each sample piece,
$D(k_x,k_y) = I(E_B=0.5$~eV$,k_y,k_y)- I(E_B=0.2$~eV$,k_y,k_y)$.
The result is shown in Fig.~\ref{Fig3}(e-h). 
The choice of these two binding energies is based on the fact that
the intensity distribution at $E_B=0.2$~eV is almost symmetric, 
and the distribution at $E_B=0.5$~eV shows a maximum asymmetry caused by the N{\'e}el vector direction. 
Additional intensity gradients are significantly reduced for the 
differential intensity distribution. 
The result indicates an asymmetric photoemission intensity distribution
in reciprocal space.
The larger intensity occurs for an in-plane moment
direction $\vec{n}\times \vec{M}$, where $\vec{M}$
denotes the remanent Permalloy magnetization and $\vec{n}$ the surface normal. 
From theory (see the bottom panel of Fig.~\ref{Fig3}), we learn that a higher intensity occurs for an in-plane moment
direction $-\vec{n}\times \vec{N}$ (extended blue area), where $\vec{N}$
is the N{\'e}el vector. 
Thus our experimental result shows that  $\vec{N}$ is antiparallel to $\vec{M}$. 
This implies that the top Mn layer in the Mn$_2$Au film has a magnetic moment antiparallel to the magnetic moments of the Permalloy.

\hspace{0.1cm}

\section{Summary}
We experimentally observe a broken symmetry of the band structure in Mn$_2$Au(001) epitaxial thin films capped with a 2~nm Permalloy film, which is caused by the homogeneous orientation of the N{\'e}el vector. 
The homogeneous orientation of the N{\'e}el vector is made possible by the exceptionally strong exchange bias field, which allows the N{\'e}el vector to be controlled by the direction of the Permalloy magnetization. XMCD-PEEM images show that a full
layer of completely uncompensated Mn moments is responsible for this exchange bias.

The broken symmetry of the band structure is most pronounced at a binding energy of 0.3-0.4~eV.
We observed a maximum asymmetry of 1\% probed by bulk-sensitive hard X-ray photoemission spectroscopy.
The N{\'e}el vector induced parity violation of the crystallographic symmetry, visible in the photoelectron momentum patterns, is consistent with 
previously published band structure calculations. 

The band asymmetry, $E(k) \neq E(-k)$, allows the
N{\'e}el vector orientation to be assigned with respect to the Permalloy magnetization. 
We find an antiparallel orientation of the 
N{\'e}el vector direction and the Permalloy magnetization,
indicating an antiparrallel orientation of the top Mn layer and the
Permalloy magnetic moments.

\begin{acknowledgments} 
Funding by the Deutsche Forschungsgemeinschaft (DFG, German Research Foundation) - TRR 173-268565370 (Projects A01, A02 and A05) 
and Scho 341/16-1 as well as BMBF 05K19UM4
is gratefully acknowledged. 
We acknowledge the Paul Scherrer Institut, Villigen, Switzerland, for the beamtime at the SIM beamline of the Swiss Light Source and we thank the SIM beamline staff for technical support.
\end{acknowledgments}

\bibliography{Mn2Au_APV,Mn2Au_Bands-extra}

\end{document}